\documentstyle[11pt,aaspp4]{article}
\newcommand \as   {$^{\prime\prime}$}  
\newcommand \cmsq           {\hbox{cm$^{-2}$}}
\newcommand \etal         {{et~al.} }
\newcommand \flam         {\hbox{ergs s$^{-1}$ cm$^{-2}$ \AA$^{-1}$}}  
\newcommand \fnu          {\hbox{ergs s$^{-1}$ cm$^{-2}$ Hz$^{-1}$}}   
\newcommand \hst          {\hbox{\it HST}}
\newcommand \kms          {\rm{\hbox{km s$^{-1}$}}}
\newcommand \lam          {$\lambda$}
\newcommand \Lya          {\hbox{Ly$\alpha$}}
\newcommand \Lyb          {\hbox{Ly$\beta$}}

\def \ul           #1{$\underline{\smash{\hbox{#1}}}$}
\newcommand \zaz          {{$z_a\kern -1.5pt \approx\kern -1.5pt z_e$}}
\newcommand \zllz         {{$z_a\kern -3pt \ll\kern -3pt z_e$}}
\newcommand \Zsun          {\hbox{Z$_{\odot}$}}
\begin{document}

\baselineskip 13.2pt
\title
{\Large\bf The Nature of Associated Absorption and the\\
UV--X-ray Connection in 3C 288.1}
\bigskip\medskip

\author
{\large Frederick W. Hamann}
\medskip
\affil
{Department of Astronomy, University of Florida, 211 Bryant Space \\ Sciences 
Center, Gainesville, FL 32611-2055 \ ({\it hamann@astro.ufl.edu})}
\medskip

\author
{\large Hagai Netzer}
\medskip
\affil
{Tel Aviv University, School of Physics and Astronomy, Tel Aviv, 69978 
Israel}

\begin{center}
and
\end{center}
\author
{\large Joseph C. Shields}
\medskip
\affil
{Department of Physics and Astronomy, Clippinger Research Labs 251B,\\ 
Ohio University, Athens, OH 45701-2979}
\bigskip
\begin{abstract} 

We discuss new {\it Hubble Space Telescope} spectroscopy of the 
radio-loud quasar, 3C~288.1. The data cover 
$\sim$590 \AA\ to $\sim$1610 \AA\ in the quasar rest frame. They 
reveal a wealth of associated absorption lines (AALs) with no 
accompanying Lyman-limit absorption. The metallic AALs 
range in ionization from \ion{C}{3} and \ion{N}{3} to \ion{Ne}{8} 
and \ion{Mg}{10}. We use these data and photoionization models 
to derive the following properties of the AAL gas: 
1) There are multiple ionization zones within the AAL region, 
spanning a factor of at least $\sim$50 in ionization 
parameter. 2) The overall ionization is consistent with 
the ``warm'' X-ray continuum 
absorbers measured in Seyfert 1 nuclei and other QSOs. 
However, 3) the column densities implied by the AALs in 3C~288.1 
are too low to produce significant bound-free absorption at any 
UV--X-ray wavelengths. Substantial X-ray absorption would 
require yet another zone, having a 
much higher ionization or a much lower velocity dispersion 
than the main AAL region. 4) The total hydrogen 
column density in the AAL gas is 
$\log N_{\rm H} ({\rm cm}^{-2})\approx 20.2$. 
5) The metallicity is roughly half solar. 
6) The AALs have deconvolved widths of $\sim$900~\kms\ and 
their centroids are consistent 
with no shift from the quasar systemic velocity (conservatively 
within $\pm$1000 \kms ). 7) 
There are no direct indicators of the absorber's location 
in our data, but 
the high ionization and high metallicity both suggest a close 
physical relationship to the quasar/host galaxy environment. 

Finally, the UV continuum shape gives no indication of a 
``blue bump'' at higher energies. 
There is a distinct break of 
unknown origin at $\sim$1030 \AA , and the decline toward higher 
energies (with spectral index $\alpha\approx -1.73$, for  
$f_{\nu}\propto \nu^{\alpha}$) is even steeper than a single 
power-law interpolation from 1030 \AA\ to soft X-rays. 

\end{abstract}
\smallskip

\keywords{Galaxies: active; Quasars: absorption lines; Quasars: general; 
Quasars: individual (3C~288.1)}
\newpage

\section{Introduction}

Associated absorption lines (AALs) in quasar spectra provide 
unique information on the kinematics, physical conditions and 
elemental abundances in the gas near quasars. AALs are defined 
empirically as having 1) relatively narrow 
profiles (less than a few hundred \kms ), 
and 2) absorption redshifts, $z_a$, near the emission 
redshift, $z_e$, (generally within 3000--5000~\kms , see   
\cite{wey79,fol86,fol88}). The first criterion distinguishes AALs  
from the class of broad absorption lines (BALs), which have velocity 
widths and maximum displacements that often exceed 10,000~\kms .  
BALs clearly form in high-velocity winds from the central engines 
(see the reviews by \cite{wey97,wey95,tur95,wey85}).   
AALs (or \zaz\ systems) can form potentially in a variety of 
locations --- from outflows near the black hole/accretion disk, 
perhaps like the BALs, to intervening gas or galaxies at large 
(cosmologically significant) distances. 
A few AAL systems are known to form in quasar ejecta (probably 
within a few pc of the energy source, \cite{h97a,bar97a,gan99}), 
while others clearly probe extended regions on host-galaxy scales 
($>$1 kpc distant, \cite{wil75,sar82,mor86,tri96,bar97b}). 
Surprisingly little else is known about the nature 
of the absorbing regions or their relationship to other quasar 
phenomena. 

One important clue is that both AALs and BALs in the UV appear to 
correlate with the presence of continuous absorption in soft X-rays 
(\cite{gre96,cre99,mat98,mat99,gal99,bra99} and references therein). 
The X-ray absorbers in AAL sources (both quasars and Seyfert 1 
galaxies) tend to have high total hydrogen column densities  
($\log N_{\rm H} ({\rm cm}^{-2})\sim 21$ to 23)  
and high ionizations (being 
dominated by absorption edges of O~{\sc vii} and O~{\sc viii} near 
0.8 keV) (\cite{rey97,geo98a}). In contrast, the AALs typically 
indicate lower column densities and 
lower levels of ionization (cf. \cite{h97}). The X-ray absorbers 
that accompany BALs have even larger total column densities 
of $\log N_{\rm H} ({\rm cm}^{-2})\ga 23$ 
(\cite{gre96,gal99}). In those objects, 
the total column densities derived from X-rays exceed prior estimates 
from the BALs by 2 or more orders of magnitude (\cite{h98a}). 
Clearly, we must consider the UV and X-ray data together  
to obtain a complete census and understanding of the 
absorbing environments. 

A key question now is the physical relationship between the 
UV and X-ray absorbers. Mathur \etal (1998 and refs. therein) 
argue that they could reasonably identify a single absorbing medium, 
while other work has shown that multiple regions 
(having different velocities, ionizations and/or column densities) 
are at least sometimes present 
(e.g. \cite{kri96,h97a,rey97,geo98b,mat99}). High-ionization UV lines, 
such as \ion{Ne}{8} \lam\lam 770,780 and \ion{Mg}{10} \lam\lam 610,625, 
can directly test the UV--X-ray relationship because their 
ionization requirements are similar to the \ion{O}{7} and 
\ion{O}{8} edges measured in X-rays. The far-UV spectra needed to 
reach the \ion{Ne}{8} and \ion{Mg}{10} lines also encompass many 
under-utilized diagnostics such as the \ion{H}{1} Lyman limit, 
the Lyman series lines, and numerous metal lines spanning a 
wide range of ionizations. Unfortunately, these features are 
difficult or impossible to measure in many sources because of 
their short wavelengths. For example, they are obscured by Galactic 
Lyman-limit absorption in low-redshift sources (e.g. in all Seyfert 
galaxies) and contaminated by the dense ``forest'' of \Lya\ 
absorbers at high redshifts. Another problem at high redshifts is 
that the dominant \ion{O}{7} and \ion{O}{8} edges are shifted out of 
the sensitive energy range of current X-ray telescopes. 

Intermediate-redshift quasars ($0.5\la z\la 1$) 
provide a unique opportunity to measure all of the key UV and 
X-ray features in the same object. We have begun a program to 
obtain UV and X-ray spectra of several such objects. 
Here we discuss new UV observations of the radio-loud quasar, 
3C~288.1. This source has both the required moderate redshift, 
$z_e\approx 0.961$ (\cite{sch68}), and a strong AAL system 
(\cite{will95}). It also has a dramatically bipolar (lobe-dominated) 
radio morphology (\cite{rei95,aku94}). 
The X-ray absorption properties of 3C~288.1 are not yet known, but it 
was weakly detected in soft X-rays with the {\it Einstein} IPC 
(from $\sim$0.16 to $\sim$3.5~keV, \cite{zam81}). Wilkes \etal (1994) 
estimate its 2-point power-law index between 2500 \AA\ and 2~keV to 
be $\alpha_{ox} = -1.51$ (where $f_{\nu}\propto \nu^{\alpha_{ox}}$). 

\section{Observations and Data Reductions}

We obtained spectra of 3C~288.1 in two observations with the 
{\it Hubble Space Telescope (HST)}. Both measurements used the 
Space Telescope Imaging Spectrograph (STIS) with a 0.2\as\ $\times$ 
52\as\ slit and the MAMA detectors. 
The first observation, on 12 January 1999, yielded 8564 s 
of on-source integration time in 3 exposures with the G230L grating. 
The usable wavelength coverage is $\sim$1645 \AA\ to 
$\sim$3160 \AA\ (observer's frame). The realized spectral resolution 
depends mostly on the line-spread function of the spectrograph for 
the quasar point source. We measure this resolution to be roughly 
5.9 \AA , or 3.7 pixels on the MAMA detector, based on the 
full widths at half minimum (FWHMs) of Galactic absorption lines. 
The velocity resolution is thus $\sim$1080 to 
$\sim$560 \kms\ from the short- to long-wavelength ends of the spectral 
coverage. The second observation, on 21 January 1999, used the G140L 
grating for a total of 14778 s in 5 exposures. 
The spectral coverage in this case is roughly 1155 to 1720 \AA\ 
at a measured resolution of $\sim$2.5 \AA\ 
($\sim$4.2 pixels on the MAMA), corresponding to 
$\sim$660 to $\sim$440 \kms\ from the short- to long-wavelength ends. 
The combined spectra provide complete wavelength coverage from 
$\sim$590 \AA\ to $\sim$1610 \AA\ in the quasar rest frame. 

We acquired flux-calibrated spectra for these observations 
(one spectrum for each exposure) 
from the Space Telescope Science Institute, based on their 
standard ``pipeline'' reductions. 
We then performed additional manipulations and measurements using 
the IRAF\footnote{IRAF is distributed by the National Optical Astronomy 
Observatories, which operates under the Association of Universities for 
Research in Astronomy in cooperative agreement with the National 
Science Foundation.} software package. In particular, 
we measured Galactic absorption lines to establish that there are 
no significant wavelength shifts between exposures taken with 
the same grating. 
We then averaged the 3 G230L spectra and the 5 G140L spectra 
using weights determined from the integration times. 
To check the absolute wavelength calibrations with each grating, 
we measured centroids for several Galactic absorption 
lines in the two averages. Based on those measurements, 
and assuming the Galactic lines are at their laboratory 
wavelengths (\cite{sch93}), we applied 
offsets of $-$1.25 \AA\ and $-$0.2 \AA\ to the mean G230L and 
G140L spectra, respectively. 

\section{Results}

Figure 1 shows the final mean spectra. 
They reveal many new AALs compared to earlier work (\cite{will95}), 
including \ion{Ne}{8} \lam\lam 770,780 and probably \ion{Mg}{10} \lam 625.  
They also show no \ion{H}{1} Lyman limit absorption related to  
the AALs (see also Fig. 3 below). 

\subsection{Absorption Lines}

Table 1 lists properties of the detected 
absorption lines, namely, the vacuum centroid wavelengths 
($\lambda_{obs}$) and equivalent widths ($W_{\lambda}$) in the 
observed frame (both in \AA ), the line identifications 
(ID), the absorption redshifts ($z_a$) for identified non-Galactic 
features, and the derived column densities as discussed in \S4.2 
below. The derived quantities use laboratory wavelengths 
and atomic data from Verner \etal (1994a). 
The strong AAL system has a nominal redshift of 
$z_a \approx 0.9627$. Table 1 includes upper limits on 
$W_{\lambda}$ for several lines 
{\it not} detected in this system. The last column in the table provides 
additional notes, including FWHMs for the strongest unblended 
lines at $z_a \approx 0.9627$.

We measure the absorption lines by first defining a pseudo-continuum 
based on smooth polynomial fits to the actual continuum and broad emission 
lines. We then use cursor functions in IRAF's {\tt splot} program to 
measure (by direct integration) significant absorption features 
relative to the fitted curve.
For unblended lines, the uncertainties in $W_{\lambda}$ are 
dominated by the subjective pseudo-continuum placement. 
We estimate the 1$\sigma$ uncertainties for these features to be 
$\la$0.1 \AA\ in the G140L data and $\la$0.15 \AA\ in G230L. 
For absorption features blended with each other, 
the uncertainties depend on the severity 
of the blend. If the blending is not severe (i.e. if there are 
still distinct absorption minima at each transition's wavelength), 
we again measure/deblend the individual lines ``by eye'' using 
cursor functions in {\tt splot}. To check our accuracy, 
we also fit some of these modestly blended lines with 
gaussian profiles (using $\chi^2$ minimization in the IRAF task 
{\tt specfit}). Figure 2 shows a blend of 4 such lines, 
including \Lyb\ and the \ion{O}{6} doublet in the $z_a\approx 0.9627$ 
system. We fit these features with one gaussian per transition 
(see dotted curves in figure). The redshifts and velocity widths 
of the \ion{O}{6} pair are forced to be identical. 
The measurements derived from these fits appear in Table 1. 
They are within 10\% of our 
estimates from manual deblending, thus confirming the viability of 
both procedures. 

Some severe blends in the $z_a \approx 0.9627$ system do not have 
distinct absorption dips corresponding to each transition. For the 
multiplets in this category, \ion{C}{4} \lam\lam 1549,1551, 
\ion{N}{5} \lam\lam 1239,1243 and \ion{N}{3} \lam\lam 685,686 (Fig. 1), 
we make no attempt at deblending and list them as single lines 
in Table 1. For other unresolved blends, we measure $W_{\lambda}$ for 
the entire blend and then divide the result among the different 
transitions. The $W_{\lambda}$ for both the entire blend and the 
individual lines are given in Table 1. In particular, 
\ion{O}{5} \lam 608 is part of an unresolved blend with 
Galactic \ion{Si}{2} \lam\lam 1190,1193. We estimate  
$W_{\lambda}$ for \ion{O}{5} \lam 608 alone by subtracting a 
prediction for the Galactic \ion{Si}{2} (1.4 \AA ), based on the 
relative strengths of the various \ion{Si}{2} lines in other quasar 
spectra (\cite{sch93}). The result is given in Table 1 without 
further correction for the possible contribution from \ion{Mg}{10} 
\lam 610. Another case involves \ion{O}{4} \lam 788 
and \ion{S}{5} \lam 786, which are blended with each other and with 
Galactic \ion{C}{4} \lam\lam 1549,1551 absorption. 
For this blend, we first subtract a \ion{C}{4} contribution 
(0.6 \AA ) derived from the Galactic \ion{C}{4}/\ion{Si}{4} ratio 
in other sources (\cite{sch93}). We then divide the remaining 
$W_{\lambda}$ between the \ion{O}{4} and \ion{S}{5} lines 
according to the ratio (3.55:1) of their oscillator strengths weighted 
by solar abundances (i.e. their ``strength'' parameters 
in \cite{ver94a}). 

Finally, we note that the measurement 
uncertainties are particularly large for the \Lya\ absorption at 
$z_a \approx 0.9627$ because it lies near a sharp peak in the 
\Lya\ emission line (Fig. 1). Our estimate of this \Lya\ absorption 
strength is therefore sensitive to the assumed emission profile. 
In \S4.3 we will show that the value of 
$W_{\lambda}$ given for this line in Table 1 is almost certainly 
too small. The 1$\sigma$ uncertainties in the \ion{O}{6}, \ion{N}{5} 
and \ion{C}{4} AALs, which also sit atop emission lines, should be 
$<$10\% based on repeated measurements with different assumed 
emission-line profiles. 

\subsection{Emission Lines}

We measure the emission lines using our fit above 
to the pseudo-continuum (\S3.1). This fit (for example 
Figure 2) interpolates across the absorption features and 
thus approximates the unabsorbed emission spectrum. We define the 
line emission relative to a subjective estimate of the ``true" 
continuum in this fitted spectrum. Direct integration then 
yields approximate rest-frame equivalent 
widths of $W_{\lambda} = 21\pm 2$ \AA\ for \ion{C}{4} \lam 1549, 
$W_{\lambda} = 12\pm 3$~ \AA\ for 
\ion{O}{6} \lam 1034, and $W_{\lambda} = 50\pm 5$~ \AA\ for the 
\Lya\ + \ion{N}{5} \lam 1240 blend. The 1$\sigma$ uncertainties 
are estimates based on multiple measurements with different plausible 
continuum placements. We estimate a 3$\sigma$ upper limit on 
the \ion{Ne}{8} \lam 774 equivalent width of very roughly 7 \AA , 
consistent with the measured strength of this feature in other 
QSOs (\cite{h98b}). 

\subsection{Continuum Shape}

Figure 3 shows the combined \hst -STIS spectra on a log-log scale 
with frequencies shifted to the quasar rest frame. There is a clear 
change in the continuum slope near 1030 \AA\ 
($\log\nu ({\rm Hz})\approx 15.46$). 
This change is illustrated in the figure 
by a broken power-law ($f_{\nu}\propto \nu^{\alpha}$) with 
$\alpha\approx -1.73$ for $\lambda\la 1030$ \AA\ and 
$\alpha\approx -0.83$ for $\lambda\ga 1030$ \AA. A break like 
this near 1030 \AA\ appears to be typical of QSOs 
(\cite{zhe97,obr88}), although 
the spectral indices derived here are less negative 
(by 0.2--0.7 dex) than the published averages for other 
radio-loud sources. 

It is interesting to compare the UV continuum fluxes in Figures 1 and 3 
with the soft X-ray measurements of 3C~288.1 (\cite{wilk94}). 
In particular, the 2-point power-law index between 2 keV and 
1030 \AA\ in the rest frame is $\alpha_{uvx}\approx -1.51$. 
Note that the X-ray measurement 
is just a $\sim$3$\sigma$ detection and that those data were 
obtained $\sim$19.5 years prior to our \hst\ observations. 
Also note that the value of $\alpha_{uvx}$ given here is 
coincidently identical to $\alpha_{ox}$ (between 2 keV and 
2500 \AA ) reported by Wills \etal (1994). They used the same 
X-ray data but a different (ground-based) measurement of the 
rest-frame UV. Evidently, the UV flux varied. The X-rays 
might have varied also. Nonetheless, the slope we measure at 
$\lambda\la 1030$ \AA\ appears, like other radio-loud quasars, 
to be significantly steeper than a single power-law 
extrapolation from 1030 \AA\ to soft X-rays (\cite{lao97}). 
The implications of this continuum shape are discussed in 
\S5.3 below. 

\subsection{UV Variability Check}

Wills \etal (1995) observed 3C~288.1 with the Faint Object Spectrograph 
(FOS) on board \hst\ in April 1993. Those spectra span observed 
wavelengths from 2225 to 3280 \AA\ at a resolution slightly higher than 
our STIS data, FWHM $\approx$ 230 \kms . Direct comparisons between the 
two data sets reveal no significant changes in the continuum (from 
1135 to 1610 \AA\ in the rest frame) or the emission or absorption 
lines (\Lya , \ion{N}{5} and \ion{C}{4}) measured in common.

\section{Properties of the Associated $z_a \approx 0.9627$ Absorber}

\subsection{Kinematics}

The line profiles in the $z_a \approx 0.9627$ system are marginally 
resolved, with typical measured FWHMs of $\sim$1100 \kms\ 
(Table 1). Simple gaussian deconvolution from the instrumental 
response profile therefore suggests that the 
intrinsic line widths are roughly 900 \kms . 
We cannot rule out the possibility that these widths 
result from a blend of many narrower (unresolved) features. However, 
the higher resolution (230 \kms ) FOS spectra of Wills et al (1995, 
see \S3.4 above) give no indication of narrower features in \Lya , 
\ion{C}{4} or \ion{N}{5}. The $\sim$900 \kms\ line widths should, 
in any case, represent the full line-of-sight velocity 
dispersion through the absorbing region. 

We do not have an accurate estimate of the absorber's velocity 
shift relative to the quasar. The emission redshift quoted in the 
literature, $z_e \approx 0.961$, 
comes from an old photographic measurement 
of \ion{C}{3}] \lam 1909 and \ion{Mg}{2} \lam 2799 (Schmidt 1968). 
Nonetheless, the difference between that $z_e$ and 
our measurement of $z_a \approx 0.9627$ is consistent with the 
apparent small blueshift of the AALs relative to the emission-line 
peaks (e.g. in \ion{C}{4}, \Lya , and \ion{O}{6}, Fig. 1). 
Interpreting this velocity shift, roughly 250 \kms , 
in terms of a radial flow is problematic, however, because the 
broad emission lines might be either blue or redshifted with 
respect to the quasar's true systemic velocity. 
For example, Marziani \etal (1996) 
measured velocity shifts up to $\pm$1000 in \ion{C}{4} relative 
to the narrow emission lines (e.g. [\ion{O}{3}] \lam\lam 4959,5007) in a 
sample of radio-loud quasars. We therefore conclude conservatively 
that the AALs in 3C~288.1 lie within 
$\pm$1000 \kms\ of the quasar's systemic velocity.

\subsection{Column Densities}

We use a curve-of-growth analysis to derive column densities for 
each of the ions detected at $z_a \approx 0.9627$. 
With the line equivalent widths from Table 1, 
the only free parameter in this analysis is the doppler $b$ value  
(where $b$ = FWHM/1.665 for gaussian line profiles). Even though 
the line profiles are not well resolved, 
the data provide several independent 
constraints on $b$ and the inferred column densities. 
In particular, 1) the deconvolved FWHMs, 
roughly 900 \kms\ (\S4.1), require $b\la 540$ \kms . 
2) The absence of an \ion{H}{1} absorption edge at 912 \AA\ 
(see Figs. 1 and 3) places a firm upper limit on the \ion{H}{1} 
column density and thus a lower limit on the Lyman-line $b$ value. 
We estimate that the optical depth 
at this edge is no more 10\%, implying an \ion{H}{1} column of 
$\log N_{\rm HI}({\rm cm}^{-2}) < 16.2$ \cmsq\ and $b\ga 190$ \kms\ 
in the Lyman lines. 3) Comparing different lines of the same ion, 
such as the \ion{H}{1} Lyman series 
and doublets like \ion{O}{6} and \ion{Ne}{8}, places firm limits on 
their $b$ values and column densities. 
The fact that these multiplets do not have observed ratios of 
$\sim$1:1 (Fig. 1 and Table 1) implies immediately that the lines 
are not dominated by very optically thick components (with small $b$). 
Finally, 4) weak lines of low-abundance elements like 
\ion{P}{5} \lam\lam 1118,1128 are not detected. If the metals have 
roughly solar relative abundances, the absence of these weak 
transitions implies that strong transitions like 
\ion{C}{4} \lam 1548 have optical depths $\la$30 (\cite{h98a}). 

Table 1 lists the column densities derived from each measured line 
in two limiting cases, $b=200$ \kms\ and $b=540$ \kms . 
Higher $b$ values would exceed the deconvolved FWHMs, 
while lower $b$ values would violate the upper limit on $N$(\ion{H}{1}) 
and yield inconsistent results for some multiplets (e.g. \ion{H}{1} 
and \ion{O}{6} \lam\lam 1032,1038). Comparing these two results in Table 
1 gives an indication of the theoretical uncertainties. 
Note that the results for $b=540$ \kms\ are all within 0.2 dex of 
the optically thin lower limits. 

We adopt an intermediate value of $b=300$ \kms\ for 
our ``best guess'' column densities. These columns are listed 
in Table 2 after averaging over all useful lines for each ion. 
Entries marked ``:'' are uncertain by as much as a factor of 
$\sim$2, while those labeled ``::'' have even larger uncertainties. 
\ion{C}{3}, \ion{Mg}{10} and \ion{S}{5} have these uncertainty 
flags because of blending problems, while \ion{S}{3} and \ion{S}{4} 
have large uncertainties because different lines yield 
substantially different column density results (Table 1). 
\ion{O}{5} is marked as uncertain because the only line measured 
for that ion, \lam 630, might be saturated. (The 
curve-of-growth analysis with $b=300$ \kms\ suggests that 
\ion{O}{5} \lam 630 has optical depths up $\sim$7, 
higher than any other measured line.) 

Note that the analysis above assumes the 
absorber fully covers the background light source(s) along our 
line(s) of sight. Partial coverage is known to occur in some AAL
systems, based on measured doublet ratios in high resolution 
spectra (\cite{wam93,pet94,h97a,bar97b,gan99}, also \S5.2 below). 
The only 2 known cases of partial 
coverage in radio-loud quasars have coverage fractions of 
$>$95\% (\cite{bar97a,h99a}). The strongest constraint 
on the coverage fraction in 3C~288.1 comes from the measured depth 
of the deepest line, \ion{O}{5} \lam 630, which requires $>$75\% 
coverage. If the coverage fraction is near 75\%, then the column 
densities listed for the deepest lines in Table 2 could be 
underestimated. On the other hand, if the coverage fraction 
is $>$95\% like the other radio-loud quasars, all of the 
column densities in Table 2 would be accurate. 

Clearly, the column densities reported 
here should be checked with higher resolution spectra. Such data 
would test for both partial coverage and narrow (presently 
unresolved) line components. Narrow lines (with low $b$ values) 
might, in principle, harbor large column densities while contributing 
little to the total equivalent widths. However, 
the line multiplet ratios and the lack of a Lyman edge already 
prohibit low $b$ values and large column densities for the 
{\it dominant} absorber in 3C~288.1 (even if there is incomplete 
coverage). The column densities in Table 2 should therefore 
apply to whatever absorber(s) control the measured equivalent widths. 
We will adopt these column densities hereafter in our discussion, 
with the understanding that they apply strictly to the dominant 
AAL absorber. 

\subsection{Ionization, Density and Radial Distance}

The detected metal lines at $z_a \approx 0.9627$ range in ionization 
from \ion{C}{3} and \ion{N}{3} to \ion{Ne}{8} and 
\ion{Mg}{10}. There are no strong transitions of higher ions, 
for example \ion{Si}{12} \lam\lam 499,521, within our wavelength 
coverage. The upper limit of 
the ionization is therefore unknown. The lower limit 
is firmly established by the absence of 
singly-ionized metals such as \ion{C}{2}, \ion{N}{2} and 
\ion{O}{2} (see Table 1). 

We assume the absorber is in photoionization equilibrium with the quasar 
radiation field  
and we use the numerical code CLOUDY (version 90.04, \cite{fer98}) 
to examine its ionization properties. We note that the column densities 
in Table 2 imply immediately that the absorber is optically 
thin in the Lyman continuum, out to at least $\sim$0.37 keV (the 
ionization threshold of \ion{Mg}{10}). Figure 4 plots calculated 
ionization fractions for \ion{H}{1} and various metal ions, M$_i$, 
in general optically 
thin clouds that are photoionized by a broken power-law spectrum. 
We use a spectral index of $\alpha = -1.7$ in the Lyman continuum 
out to 1~keV, based mainly on our measurement of $\alpha = -1.73$ 
for $\lambda\la 1030$ \AA\ (\S3.3). At higher energies 
we adopt the slope $\alpha_x = -0.9$, based on X-ray 
observations of similar objects (\cite{lao97,geo98a}). There is 
considerable uncertainty about the true spectral shape of 3C~288.1 
(and other quasars) in the extreme UV and soft X-rays. 
The uncertain slope is particularly 
important when comparing ions/lines with very different ionization 
energies. However, these uncertainties are not important to our 
main conclusions below.   

The ion fractions in Figure 4 are plotted 
for a range of ionization parameters, 
$U$ --- defined here as the dimensionless ratio of hydrogen-ionizing 
photon to hydrogen particle densities\footnote{For comparison, we note 
that $\log U_x = \log U - 1.47$ for this continuum shape, where 
$U_x$ is the ionization parameter defined over the photon energies 
0.1 to 10 keV (\cite{net96}).} (see \cite{fer98}). 
The ionization fractions are not 
sensitve to either the metal abundances or the 
space density (for a given $U$). They also do not depend on the 
column densities used in the 
calculations, as long as the gas remains optically thin in the 
ionizing continuum (see \cite{net96,h95,h97} for more 
discussion and calculations using other parameters).

We estimate $U$ in the 3C~288.1 absorber 
by comparing the ion fractions $f$(M$_i$) in Figure 4 to 
various column density ratios from Table 2. 
Each column density ratio yields an independent 
estimate. Some of the $U$ values are illustrated in Figure 4 by 
bold vertical lines that connect the $f$(M$_i$) curves used for that 
estimate. For example, the measured \ion{N}{3}/\ion{N}{4} column 
density ratio is $-$0.6 dex, implying a moderate ionization with 
$\log U\approx -1.7$. 
The \ion{C}{3}/\ion{C}{4} and \ion{O}{3}/\ion{O}{4} ratios 
suggest similar $U$. (Although included in Figure 4, 
we consider the $U$ estimates for sulfur to be 
unreliable because of the uncertainties in their column 
densities, \S4.2.) By making the reasonable assumption 
that O, Ne and Mg have roughly solar relative abundances, we 
use the \ion{O}{6}/\ion{Ne}{8} and \ion{Mg}{10}/\ion{Ne}{8} 
column densities to infer much larger values of 
$\log U\approx 0.0$ in the region where those lines form. 
The ratios of intermediate ions like 
\ion{N}{4}/\ion{N}{5} and \ion{O}{5}/\ion{O}{6} 
indicate intermediate $U$ values\footnote{Note that these simple 
$U$ estimates do not correct for multiple zones possibly contributing 
to the measured column densities. Such corrections would require 
and explicit (but ad hoc) model.}. 

The differences in these derived $U$ values are well above the 
uncertainties and inconsistent with a single zone absorber. 
For a given space 
density, $n_H$, and distance, $R$, from the ionizing 
continuum source, optically thin clouds should have the same level 
of ionization throughout (e.g. there can be no gradient 
in the ionization due to the absorber's own 
opacity). The optically thin clouds in 3C~288.1 must therefore 
occupy a range of densities or distances.  From the 
definition of $U$ (where $U\propto n_H^{-1}\, R^{-2}$), 
and the difference of $\Delta\log U \ga 1.7$ between the ``high'' 
and ``low''-ionization regions,  
we infer a factor of $\ga$50 range in the space density, 
a factor of $\ga$7 range in the distance, or some equivalent 
combination of different $n_H$ and $R$ values. 

\subsection{Elemental Abundances}

The relative abundance of any two elements $a$ and $b$ can be derived 
from the following expression, 
\begin{equation}
\left[{a\over b}\right] \ = \ \
\log\left({{N(a_i)}\over{N(b_j)}}\right) \ +\
\log\left({{f(b_j)}\over{f(a_i)}}\right) \ +\ 
\log\left({{b}\over{a}}\right)_{\odot}
\end{equation}
where $(b/a)_{\odot}$ is the solar abundance ratio (\cite{gre89}), 
and $N$ and $f$ are respectively the column densities and ionization 
fractions of element $a$ in ion state $i$, etc. With suitable  
ionization corrections, $f(b_j)/f(a_i)$ from Figure 4, we can 
trivially derive abundance ratios from the column densities in Table 2. 
For example, we estimate average values of [C/O]~$\approx -0.5$ 
and [N/O]~$\approx +0.1$ from the ratios \ion{C}{3}/\ion{O}{3}, 
\ion{C}{4}/\ion{O}{4}, \ion{N}{3}/\ion{O}{3} and 
\ion{N}{4}/\ion{O}{4}. The theoretical uncertainties 
in these results (e.g. for different continuum shapes) should be 
small, $\la$0.1 dex (at 1$\sigma$ or $\sim$60\% confidence),  
because we are comparing similar ions (\cite{h97}). 
The observational uncertainties are larger but more 
difficult to assess; we estimate that they are 
$<$0.2 dex in these averaged ratios (\S4.2 and Table 2). We derive 
the overall metallicity by assuming the \ion{H}{1} absorber resides 
mainly with the doubly- and triply-ionized metals 
at $\log U\approx -1.8$ to $-$1.6 (Fig. 4, \S4.3). 
Plugging the appropriate ion fractions into Equation 1 then 
implies [C/H]~$\approx -0.7$, [N/H]~$\approx -0.1$, 
and [O/H]~$\approx -0.3$. 
The theoretical uncertainties in this case are larger, 
perhaps up to 0.3--0.4 dex (\cite{h97}), because the ions being 
compared have significantly different ionization energies. 

If we had assumed that most of the 
\ion{H}{1} coexists with the high ions at $\log U\approx 0.0$ 
(Fig. 4, \S4.3), we would have inferred much lower metallicities 
of [M/H]~$\approx -1.7$ to $-$1.4 for the metals O, Ne and Mg. 
However, these low metallicities would lead 
to a contradiction for the lower ions in Equation 1 (for example, 
by predicting too 
much \ion{H}{1} for the measured amounts of the metal ions). 
The higher metallicities derived from the lower ions, and 
our original assumption 
that the \ion{H}{1} resides mainly in the lower $U$ gas, must 
therefore be correct. 

Nonetheless, there are still uncertainties related to 
the absorber's complexity; the gas does not have a 
single $U$ value and we do not know how much of the 
\ion{H}{1} resides co-spatially with each metal ion. We 
therefore estimate lower limits on the metal-to-hydrogen 
ratios. Hamann (1997) showed that the ionization corrections 
$f$(\ion{H}{1})/$f$(M$_i$) all have minimum values at some 
particular $U$ (see also \cite{ber86}). 
If the actual absorber has zones with different 
$U$ contributing to the lines, it can only mean 
that the true ionization corrections are larger. Therefore, the 
minimum ionization corrections yield robust minimum values 
of [M/H]. Hamann \& Ferland (1999) plot minimum ionization 
corrections for optically thin clouds photoionized by different 
power-law spectra (their Fig. 11). For the column densities in 
Table 2 and a spectrum similar to that adopted in 
\S4.3, the Hamann \& Ferland (1999) calculations imply 
lower limits of [C/H]~$\ga -0.7$, [N/H]~$\ga -0.2$, and 
[O/H]~$\ga -0.5$ for the $z_a\approx 0.9627$ absorber in 3C~288.1. 

We conclude that the overall metallicity (dominated by O/H) 
is roughly 1/2 \Zsun , with a firm lower limit near 1/3 \Zsun . 

\subsection{Total Column Densities and Predicted X-Ray Absorption}

We just argued that the \ion{H}{1} column density is contributed 
mostly by a low-ionization region, where $\log U\approx -1.8$ to 
$-$1.6 and $f$(\ion{H}{1})~$\approx -3.8$ to $-$3.5 (Fig. 4). 
The measured value of the \ion{H}{1} column density 
(Table 2) therefore implies a total hydrogen column 
of $\log N_{\rm H} ({\rm cm}^{-2})\approx 19.5$ in that region. 
If the metal abundances are approximately 1/2 \Zsun\ (\S4.4), 
we can estimate $N_{\rm H}$ in the 
high-ionization gas from the column densities (Table 2) 
and ionization fractions (Fig. 4) of \ion{Ne}{8} and \ion{Mg}{10}. 
We find $\log N_{\rm H} ({\rm cm}^{-2})\approx 20.2$ for that region. 

These results make specific predictions for the X-ray absorption 
that should accompany the UV AALs. Explicit photoionization 
calculations, using the column densities, ionizations and abundances 
quoted above, imply that the continuum optical depths should be 
$<$0.016 at 0.2 keV and $<$0.003 at 2.0 keV (in absorber's rest 
frame). The deepest X-ray absorption, due mainly to the combined 
\ion{O}{7} and \ion{O}{8} edges near 0.8 keV, should be $\la$3\% 
below the continuum. We therefore 
expect no significant X-ray absorption by the AAL gas in 3C~288.1. 
This prediction is not sensitive to the uncertain continuum shape 
or any other assumptions in the calculations. The only 
possibility for strong X-ray 
absorption is if that absorber contributes 
negligibly to the UV lines. Such an absorber would need to have 
either a much lower $b$ value or much higher ionization than 
we infer from the AALs. 

\section{Discussion}

\subsection{The UV--X-Ray Absorber Connection}

The main results of this paper are 1) the detection of the 
high-ionization AALs \ion{Ne}{8} \lam\lam 770,780 and \ion{Mg}{10} 
\lam 625, and 2) the prediction that the AAL gas will not 
produce significant bound-free absorption in X-rays (\S4.5). 
Simple 1-zone models of strong UV line {\it and} X-ray continuum 
absorption cannot apply to this object. In fact, the variety of 
UV AALs alone requires multiple absorbing zones with different 
levels of ionization (\S4.3). Strong X-ray absorption would require 
yet another zone, having a high-column density of gas with either 
much higher ionization or a much lower velocity dispersion 
($b$ value) than the main AAL absorber (\S4.5). 

To our knowledge, 3C~288.1 is now the fourth quasar for which the 
\ion{Ne}{8} AALs (and in this case also 
\ion{Mg}{10}) have been measured. The other 
quasars are UM~675 at redshift $z_e = 2.15$ (\cite{h95,h97a}), 
HS~1700+6416 at $z_e=2.713$ (\cite{pet96}), 
and J2233$-$606 at $z_e=2.24$ (\cite{pet99}). There is another 
object with well-measured \ion{Ne}{8}, \ion{Mg}{10} and even 
\ion{Si}{12} absorption (SBS~1542+541 at $z_e= 2.36$, 
\cite{tel98}), but the lines in that case appear 
more like BALs than AALs (blueshifted by $\sim$11,500 \kms\ 
and having FWHM~$\approx$~2500 \kms ). 
It is not yet clear how common these high-ionization lines 
are in AAL or BAL systems, but we know of
no cases where high-quality, short-wavelength spectra clearly 
rule out their presence. The AAL column densities measured for 
UM~675 and J2233$-$606 imply that their UV lines also 
form in multi-zone regions with no significant X-ray 
opacity. The total absorbing columns inferred from their AALs 
are similar to 3C~288.1, $\log N_{\rm H} ({\rm cm}^{-2})\la 20$ for 
solar abundances. Reliable column densities 
are not available for the UV absorbers in 
HS~1700+6416 and SBS~1542+541, although Telfer 
\etal (1998) estimate $\log N_{\rm H} ({\rm cm}^{-2})\ga 21.5$ 
for the latter BAL-like source. 

Existing X-ray data are sparse for these 5 quasars with known 
\ion{Ne}{8} absorption. The UV--X-ray spectral index of 3C~288.1, 
$\alpha_{ox}\approx -1.51$, implies a slightly above 
average X-ray/UV flux ratio for quasars of similar redshift 
and luminosity (\cite{wilk94}). 
While not a strong constraint on the X-ray absorption, this 
result is at least consistent with our prediction for negligible 
absorption based on the AALs alone (\cite{bra99}). 
Better data for HS~1700+6416 
(\cite{yua98}) indicate a modest X-ray absorbing column of  
$\log N_{\rm H} ({\rm cm}^{-2})\approx 20.5$. In contrast, the 
BAL-like source SBS~1542+541 appears to be heavily absorbed 
in X-rays, based on its weak X-ray flux (\cite{tel98,yua98}). 

Several recent studies describe the correlated appearance of 
UV and X-ray absorbers in quasars and active galaxies (\S1). 
Brandt \etal (1999) have gone 
a step further by suggesting that the strengths of the 
UV and X-ray features also correlate. Nonetheless, the 
physical relationship between the absorbers remains unknown. 
AAL regions can clearly have a variety of properties 
(e.g. ionization and column density) that 
are not always conducive to X-ray continuum absorption 
(see also \cite{kri96,h97,h97a,mat99}). Fundamentally, 
the column densities needed for strong AALs can be much less 
than those required for significant bound-free opacity. 
As a result, even those AAL regions with high, 
warm absorber-like ionizations (as in 3C~288.1, etc.) 
need not produce measurable X-ray absorption. Conversely,  
the high ionizations and high column densities derived for 
X-ray warm absorbers (\cite{rey97,geo98a}) will not {\it necessarily} 
produce strong AALs, especially in the lower ions if the velocity 
dispersion ($b$ value) is small\footnote{For 
example, if the X-ray absorber has solar abundances and 
``nominal'' parameters, such as $\log N_{\rm H} ({\rm cm}^{-2})\approx 22$ 
and $\log U\approx 0.5$ (\cite{geo98a}), then the column density in 
C~IV will be $\log N ({\rm cm}^{-2})\la 14.1$ (using ion 
fractions from Figure 4). 
The resulting line strengths depend keenly on the velocity dispersion. 
If the velocities are strictly thermal, then $b\approx 4.5$ \kms\ 
for carbon in a 15000 K gas and the equivalent width of 
C~IV \lam 1548 would be just $\sim$0.08 \AA\ in the rest frame. 
At larger velocity dispersions, say $b=20$ \kms , this line's strength 
would be $\sim$0.24 \AA , etc., up to $\sim$0.5 \AA\ in the 
high $b$ (optically thin) limit.}. 
The two absorbing regions might be physically related in general, 
but they need not be identical. 
One specific possibility, proposed for 
quasar BALs (\cite{mur95}), is that the X-ray absorption 
occurs at the base of an accretion-disk outflow while the UV 
lines form largely in the accelerated gas farther out. 

Studies that encompass both the high-ionization UV lines and 
the high-ionization X-ray edges (e.g. \ion{O}{7} and \ion{O}{8}) 
are needed to test this physical relationship further. 
Intermediate-redshift quasars like 3C288.1 are prime 
targets for this work (\S1). However, a special concern with quasar 
AALs is that they can form potentially in a variety of locations. 
The origin and physical nature of the 
AAL gas, as well as its relationship to the X-ray features, 
are thus intimately tied to the question of the absorber's location. 

\subsection{Where is the Absorber in 3C~288.1?}

Recent studies indicate that AALs have generally high 
metallicities (near or above the solar value, 
e.g. \cite{pet94,pet99,h99b}), and strengths that correlate 
with the quasar's radio properties and perhaps luminosity 
(\cite{and87,fol88,ald94,will95,bak96,pbar97,ric99,bra99}). 
These results suggest that AALs are often physically related 
to quasars. Further evidence for a close relationship 
has come from spectroscopic indicators, such as 
1) time-variable line strengths, 2) well-resolved AAL profiles that 
are smooth and broad compared to thermal line widths, and 3) multiplet 
ratios that imply partial coverage of the background light source(s) 
(e.g. \cite{bar97a,bar97b,h97a,gan99}). The link between these 
properties and the near-quasar environment is strengthened by the 
fact that they are common among BALs (\cite{bar93,bar94,h98a,ara99}). 
AALs with these properties are likely to form in outflows from 
the central engines, at radii of tens of pc or possibly much less 
(e.g. \cite{h97a}). 
It should be a high priority to apply these spectroscopic tests of 
an ``intrinsic'' origin to the AALs in 3C~288.1 and other sources. 
The existing data for 3C~288.1 provide only indirect evidence 
for intrinsic absorption. In particular: 

1) The ionization (\S4.3) and metallicity (\S4.4) 
in 3C~288.1 are probably both too high for an 
absorber at large (cosmologically significant) distances from the 
QSO (\cite{ver94b,h99b}). The high metallicity suggests that the 
AAL gas resides (or originated) within the 
quasar's host galaxy, while the high ionization suggests a strong 
influence of the quasar's intense 
radiation field. If the gas is photoionized by the quasar 
spectrum, the relationship between the gas' 
density, ionization parameter and distance from 3C~288.1 is 
$R\approx 60\; (1/U)^{1/2}\,(10^4\,{\rm cm}^{-3}/n_H)^{1/2}$ pc  
(based on the measured flux in Fig. 1, the ionizing spectral shape 
used in \S4.3, and a cosmology with 
$H_o = 65$ km s$^{-1}$ Mpc$^{-1}$ and $q_o = 0.2$). 

2) Three of the 5 absorption-line systems with measured \ion{Ne}{8} 
(UM~675, J2233$-$606 and SBS1542+531 discussed in \S5.1) are 
known to have at least one of the spectral indicators of 
an intrinsic origin listed above (\cite{h95,h97a,tel98,pet99}). 
These results support the argument in point (1) above, 
that high-ionization 
lines like \ion{Ne}{8} and \ion{Mg}{10} are additional signatures 
of intrinsic absorption. 

3) Weaker statistical evidence comes from the 
bipolar radio morphology of 3C~288.1 (\cite{rei95,aku94}). 
Lobe-dominated radio structures are known to correlate 
with the presence of strong AALs (\cite{will95,pbar97}). 
They are also indicative of an edge-on 
orientation for the central quasar/accretion disk. 
Strong AALs (as in 3C~288.1) should therefore 
tend to form not only near the quasar, but near the plane 
of its accretion disk or extended torus (\cite{will95,pbar97}). 
The situation might be analogous to
BAL outflows, where disk-like geometries are favored 
by polarization observations (\cite{goo95,hin95,coh95}) 
and by some theoretical 
models of disk-driven outflows (e.g. \cite{mur95,dek97}). 

We conclude conservatively that the AALs 
in 3C~288.1 are physically related to the quasar, probably forming
well within the radius of the quasar's host galaxy. 

\subsection{The Spectral Energy Distribution}

The shape of the ionizing continuum is a fundamental question in 
studies of quasars and active galaxies. 
A simple analysis of the broad emission-line equivalent widths 
suggests that there must be a large ``blue bump,'' peaking around 
2--5 Rydberg and contributing more than 50\% of the bolometric 
luminosity (e.g. Mathews \& Ferland 1987, Netzer 1990). 
AGN spectra cannot be measured directly at these energies,  
but recent far-UV and soft X-ray observations give no evidence for 
the predicted big blue bump (Zheng et al. 1996 and Laor et al. 1997). 
Netzer (1985) showed that there is a serious 
energy budget problem in photoionization models of the 
emission-line gas if a strong blue 
bump is not included. In particular, the line strengths are severely 
underpredicted by the models. This theoretical problem depends 
partly on the global covering factor of the emitting gas. The 
rarity of rest-frame Lyman-limit absorption in quasars implies 
that we rarely (if ever) view them through the line-emitting 
gas; thus the covering factors are expected to be low --- of order 
10\% (e.g. Antonnucci \etal 1989, Koratkar \etal 1992). 
Korista et al. (1997) recently confirmed that these low covering 
factors, together with the bump-less ionizing continuum 
inferred from the Zheng et al. and Laor et al. data, 
underpredict the emission line strengths by factors of several. 
They estimate, for example, that excessively large covering factors 
of 56\% --75\% would be needed to match the typical HeII \lam 
1640 line strength. One possible solution, proposed by Korista et al., 
is that the broad emission line regions might not ``see'' 
the same continuum shape we do. 

Currently this theoretical problem is unresolved. The 
spectrum of 3C~288.1 presented here (\S3.3) is consistent with the 
other observations, providing no evidence for a substantial blue bump 
at higher (unobserved) energies. 

\section{Summary and Conclusions}

Comparative 
UV and X-ray studies are beginning to yield new insights into 
the nature of AGN absorbing environments. However, the 
the physical relationship of the UV and X-ray absorbers 
remains unclear. 
3C~288.1 provides another counterexample to simple models that 
would attribute all of the UV and X-ray features to identically the 
same gas (see \S1 for references). Realistic models must include 
the increasing levels of complexity implied by new and better data. 
Intermediate-redshift quasars will be an important source of 
new data because they allow us to measure a wide range of 
under-utilized features in the rest-frame far UV --- 
for example  the \ion{H}{1} Lyman limit and 
high-ionization AALs like \ion{Ne}{8} \lam\lam 770,780 
and \ion{Mg}{10} \lam\lam 610,625. Our analysis of these features  
in 3C~288.1 shows that multiple absorbing regions (spanning a range 
of ionizations, densities and/or distances from the quasar) 
contribute to the AAL spectrum (\S4.3). Moreover, the column 
densities inferred from the AALs are too small to produce significant 
bound-free opacity at any UV through X-ray wavelengths (\S4.5). 
There will be no significant X-ray absorption 
in this object unless it occurs in yet another region ---
having a much higher ionization or a 
much lower velocity dispersion than the dominant AAL absorber. 
We are now pursuing X-ray observations of 3C~288.1 to test 
these predictions. 

One essential ingredient for any model of these regions 
is the location of the absorbing gas. Indirect evidence suggests 
that the AALs in 3C~288.1 form close to the quasar, 
probably well within the radius of the quasar's host galaxy 
(\S5.2). Unfortunately, that weak constraint on the 
location allows for a variety of possible absorption sites, 
e.g. interstellar gas in the extended host galaxy, a 
dense torus surrounding the active nucleus, or an outflow from the 
central engine/accretion disk. The AAL kinematics indicate that 
the absorber is clearly {\it not} part of a high-velocity wind 
like the BALs; the lines are only $\sim$900 \kms\ wide and 
their centroids are within $\pm$1000 \kms\ 
of the quasar rest frame (\S4.1). 
Better constraints on the location and kinematics 
will require higher resolution spectra and, ideally, repeated 
observations at both UV and X-ray wavelengths (to test 
for variable absorption, cf. \cite{h97a,bar97a,geo98b}). 
Those data would also provide more complete and more reliable 
estimates of the ionizations, kinematics, abundances and 
column densities in the different absorbing zones.

\bigskip

\acknowledgments
We are grateful to Gerald Kriss and others 
at the Space Telescope Science Institute for their generous 
help with the observations and data processing. We also thank the 
referee, Smita Mathur, for helpful comments. Financial support for 
this work was provided by NASA through grant GO-07356-96A 
from the Space Telescope Science Institute, which is operated by 
the Association for Research in Astronomy, Inc., under NASA 
contract NAS5-26555. FH also acknowledges 
support from NASA through grant number NAG5-3234 in their  
Long Term Space Astrophysics program. 

%
%
%
\singlespace
\begin{deluxetable}{cclcccl}
\tablewidth{0pt}
\tablecaption{Absorption Line Data}
\tablehead{\colhead{}& \colhead{}& \colhead{}& \colhead{}& 
\multicolumn{2}{c}{--- log $N$(cm$^{-2}$) ---}\\
\colhead{\ \ $\lambda_{obs}$\ \ }& \colhead{\ \ $W_{\lambda}$\ \ }& 
\colhead{ID}& \colhead{$z_a$}& \colhead{$b=200$}& \colhead{$b=540$}& 
\colhead{Notes\tablenotemark{a}\ \ }}
\startdata
1180.0& 0.41&  \nodata & \nodata & \nodata & \nodata & \nl
1193.5&  3.10&   \nodata & \nodata & \nodata & \nodata & ubl\nl
\nodata &      1.70&   \ion{O}{4} 608& \nodata &  15.9&  15.7&   ubl-i \nl
1200.4&  1.06&  \ion{N}{1} 1200& \nodata & \nodata & \nodata & bl,mult,Gal\nl
1206.8&  1.07&  \ion{S}{3} 1206& \nodata & \nodata & \nodata & bl,Gal\nl
1226.3&  0.27&   \ion{Mg}{10} 625& 0.9622&  15.0&  15.0&   bl\nl
1236.0&  3.78&   \ion{O}{5} 630& 0.9628&  16.7&  15.2&  FWHM=1080\nl
1249.1&  0.34& \nodata & \nodata & \nodata & \nodata & \nl
1260.2&  1.14& \ion{Si}{2} 1260& \nodata & \nodata & \nodata & Gal\nl
(1265.2)& \llap{$<$}0.25&   \ion{N}{2} 645& \nodata &  \llap{$<$}14.2&  \llap{$<$}14.2&   up\nl
1271.1&  0.19& \nodata & \nodata & \nodata & \nodata & \nl
1290.0&  0.51&   \ion{S}{4} 657& 0.9625&  13.8&  13.8&         \nl
1303.1&  2.04&   \ion{O}{1}  1302& \nodata & \nodata & \nodata & ubl,Gal\nl
1330.3&  0.40&   \ion{S}{3} 678& 0.9628&  13.5&  13.5&         \nl
1335.0&  1.45&   \ion{C}{2}  1335&  \nodata & \nodata & \nodata & Gal\nl
1344.7&  0.84&   \ion{N}{3} 685& 0.9622& 14.5&  14.4&   mult\nl
(1348.5)& \llap{$<$}0.25&   \ion{C}{2} 687& \nodata &  \llap{$<$}14.0&  \llap{$<$}14.0&   up\nl
(1374.4)& \llap{$<$}0.25&   \ion{Ar}{8} 700& \nodata &  \llap{$<$}13.9&  \llap{$<$}13.9&   up\nl
1377.7&  0.98&   \ion{O}{3} 702& 0.9615& 15.0&  15.0&         \nl
1393.9&  0.43&   \ion{Si}{4} 1394& \nodata & \nodata & \nodata & Gal\nl
1399.5&  0.71& \nodata & \nodata & \nodata & \nodata & \nl
1403.0&  0.95&  \ion{Si}{4} 1403&  \nodata & \nodata & \nodata & bl?,Gal\nl
1412.2&  0.42&  \nodata & \nodata & \nodata & \nodata & \nl
1421.9&  0.45&   \ion{S}{3} 724& 0.9631& 14.2&  14.2&         \nl
1456.9&  0.52&  \nodata & \nodata & \nodata & \nodata & \nl
1463.3&  0.63&   \ion{S}{4} 745& 0.9645& 14.5&  14.4&   bl \nl
1468.5&  0.85&   \ion{S}{4} 748& 0.9622& 14.3&  14.3&   bl \nl
(1481.7)& \llap{$<$}0.25&   \ion{Ar}{6} 755& \nodata &  \llap{$<$}14.6&  \llap{$<$}14.6&   up\nl
1501.5&  3.17&   \ion{N}{4} 765& 0.9624& 15.4&  14.8& FWHM=1140\nl
1512.1&  2.21&   \ion{Ne}{8} 770& 0.9629& 15.6&  15.4& FWHM=1060\nl
1520.7&  0.31&  \nodata & \nodata & \nodata & \nodata & \nl
1527.5&  1.10&  \ion{Si}{2} 1527& \nodata & \nodata & \nodata & Gal\nl
1532.8&  1.17&   \ion{Ne}{8} 780& 0.9643& 15.5&  15.4&         \nl
1536.6&  0.31&  \nodata & \nodata & \nodata & \nodata & \nl
1545.7&  4.31&  \nodata & \nodata & \nodata & \nodata & ubl\nl
\nodata &      0.82&   \ion{S}{5} 786& \nodata &  13.8&  13.8&   ubl-i\nl
\nodata &      2.89&   \ion{O}{4} 788& \nodata &  15.9&  15.5&   ubl-i\nl
\nodata &      0.60&   \ion{C}{4} 1549& \nodata & \nodata & \nodata & ubl-i,Gal\nl
(1589.1)& \llap{$<$}0.30&   \ion{S}{4} 810& \nodata &  \llap{$<$}14.4&  \llap{$<$}14.4&   up\nl
1608.1&  0.56&  \ion{Fe}{2} 1608& \nodata & \nodata & \nodata & Gal\nl
1635.2&  0.71&   \ion{O}{3} 833& 0.9632& 14.8&  14.8&   ID?\nl
(1637.8)& \llap{$<$}0.50&   \ion{O}{2} 834& \nodata &  \llap{$<$}14.5&  \llap{$<$}14.5&   up\nl
1663.0& 1.48& \nodata & \nodata & \nodata & \nodata & \nl
1671.7& 1.07&   \ion{Al}{2} 1671& \nodata & \nodata & \nodata & Gal\nl
1760.0& 1.15& \nodata & \nodata & \nodata & \nodata & \nl
1768.6& 0.92& \nodata & \nodata & \nodata & \nodata & \nl
(1774.2)& \llap{$<$}0.50&   \ion{C}{2} 904& \nodata &  \llap{$<$}14.1&  \llap{$<$}14.0&   up\nl
(1797.1)& \llap{$<$}0.50&   \ion{N}{2} 916& \nodata &  \llap{$<$}14.4&  \llap{$<$}14.3&   up\nl
1830.3& 0.94& \nodata & \nodata & \nodata & \nodata & \nl
(1831.9)& \llap{$<$}0.50&   \ion{S}{6} 933& \nodata &  \llap{$<$}13.9&  \llap{$<$}13.9&   \nl
(1840.6)& \llap{$<$}0.50&   Ly$\epsilon$ 938& \nodata &  \llap{$<$}15.7&  \llap{$<$}15.6&   up\nl
1863.5&  1.49&   Ly$\delta$ 950& 0.9621&  16.0&  15.9&         \nl
1895.8&  2.00& Ly$\gamma$ 973& 0.9493& \nodata & \nodata & \nl
1909.7&  2.35&   Ly$\gamma$ 973& 0.9636& 15.9&  15.8&   bl \nl
1917.5&  1.87&   \ion{C}{3} 977& 0.9626& 14.3&  14.2&   bl \nl
1943.1&  0.36&   \ion{N}{3} 990& 0.9631& 14.3&  14.2&      \nl
1962.0&  0.64& \nodata & \nodata & \nodata & \nodata & \nl
1997.0&  1.31&  Ly$\beta$ 1026& 0.9469& \nodata & \nl
2012.8&  4.13&  Ly$\beta$ 1026& 0.9624& 16.1&  15.6& FWHM=1190\nl
2025.9&  5.43&   \ion{O}{6} 1032& 0.9632& 16.5&  15.5&   FWHM=1140\nl
2037.0&  4.49&   \ion{O}{6} 1038& 0.9632& 16.3&  15.7&   FWHM=1140\nl
2084.2&  0.67&   \ion{S}{4} 1063& 0.9613& 14.9&  14.9&   ID?\nl
(2127.6)& \llap{$<$}0.40&   \ion{N}{2} 1084& \nodata &  \llap{$<$}14.3&  \llap{$<$}14.3&   up\nl
2144.4&  2.00& \nodata & \nodata & \nodata & \nodata & \nl
2166.5&  1.24&  \nodata & \nodata & \nodata & \nodata & \nl
(2194.3)& \llap{$<$}0.40&   \ion{P}{5} 1118& \nodata &  \llap{$<$}13.6&  \llap{$<$}13.6&   up\nl
2245.8&  1.17& \nodata & \nodata & \nodata & \nodata & \nl
2321.1&  1.43&  \nodata & \nodata & \nodata & \nodata & \nl
2344.2&  1.90&   \ion{Fe}{2} 2344& \nodata & \nodata & \nodata & Gal\nl
2366.5&  4.56&  Ly$\alpha$ 1216& 0.9466& \nodata & \nodata & \nl
2385.6&  4.40&    Ly$\alpha$ 1216& 0.9627& 15.1&  14.7&         \nl
2434.7&  7.21&   \ion{N}{5} 1240& 0.9630& 15.4&  15.2&   mult \nl
(2473.8)& \llap{$<$}0.40&   \ion{Si}{2} 1260& \nodata &  \llap{$<$}13.1&  \llap{$<$}13.1&   up\nl
2586.2&  0.85&  \ion{Fe}{2} 2587& \nodata & \nodata & \nodata & Gal\nl
2599.8&  1.24&  \ion{Fe}{2} 2600& \nodata & \nodata & \nodata & Gal\nl
(2619.3)& \llap{$<$}0.40&   \ion{C}{2} 1335& \nodata &  \llap{$<$}14.0&  \llap{$<$}14.0&   up\nl
(2735.5)& \llap{$<$}0.40&   \ion{Si}{4} 1394& \nodata &  \llap{$<$}13.4&  \llap{$<$}13.4&   up\nl
2796.1&  1.93&  \ion{Mg}{2}  2796& \nodata & \nodata & \nodata & Gal\nl
2803.9&  1.39&  \ion{Mg}{2}  2804& \nodata & \nodata & \nodata & Gal\nl
2854.2&  0.50&  \ion{Mg}{1}   2853&  \nodata & \nodata & \nodata & Gal\nl
3040.1&  7.13&   \ion{C}{4} 1549& 0.9625& 15.0&  14.8&   mult \nl
\enddata
\tablenotetext{a}{Abbreviations have the following meanings:
bl = blended line measured separately, ubl = unresolved blend, ubl-i = 
unresolved blend with individual $W_{\lambda}$ estimated, 
FWHM = full width at half minimum in km s$^{-1}$ , Gal = Galactic line, 
ID? = identification uncertain, mult = unresolved multiplet listed as 
one line, up = 3$\sigma$ upper limits at $\lambda_{obs}$ 
defined by $z_a = 0.9627$.}
\end{deluxetable}

\begin{deluxetable}{lc}
\tablewidth{0pt}
\tablecaption{Column Densities}
\tablehead{\colhead{} & \colhead{log $N$(cm$^{-2}$)}\\
\colhead{Ion\ \ }& \colhead{$b=300$}}
\startdata
\ion{H}{1}& 15.8\nl
\noalign{\vskip 5pt}
\ion{C}{2}& \llap{$<$}14.0\nl
\ion{C}{3}& 14.3\rlap{:}\nl
\ion{C}{4}& 14.9\nl
\noalign{\vskip 5pt}
\ion{N}{2}& \llap{$<$}14.2\nl
\ion{N}{3}& 14.4\nl
\ion{N}{4}& 15.0\nl
\ion{N}{5}& 15.2\nl
\noalign{\vskip 5pt}
\ion{O}{2}& \llap{$<$}14.5\nl
\ion{O}{3}& 14.9\nl
\ion{O}{4}& 15.7\nl
\ion{O}{5}& 15.6\rlap{:}\nl
\ion{O}{6}& 15.8\nl
\noalign{\vskip 5pt}
\ion{Ne}{8}& 15.4\nl
\noalign{\vskip 5pt}
\ion{Mg}{10}& 15.0\rlap{:}\nl
\noalign{\vskip 5pt}
\ion{Si}{2}& \llap{$<$}13.1\nl
\ion{Si}{4}& \llap{$<$}13.4\nl
\noalign{\vskip 5pt}
\ion{P}{5}& \llap{$<$}13.6\nl
\noalign{\vskip 5pt}
\ion{S}{2}& \llap{$<$}13.5\nl
\ion{S}{3}& 14.0\rlap{::}\nl
\ion{S}{4}& 14.2\rlap{::}\nl
\ion{S}{5}& 13.8\rlap{::}\nl
\ion{S}{6}& \llap{$<$}13.9\nl
\enddata
\end{deluxetable}


\newpage

\centerline{FIGURE CAPTIONS}
\bigskip

\noindent\ul{Fig. 1.} --- \hst\ STIS spectra of 3C~288.1 at both the 
observed and rest-frame wavelengths (where rest is defined by the 
emission redshift $z_e=0.961$). The upper and lower panels show 
spectra from the G140L and G230L gratings, respectively. 
The Flux has units $10^{-15}$~\flam . 
Absorption lines in the main AAL system ($z_a \approx 0.9627$) 
are labeled across the top. Galactic lines and a weaker \Lya\ system 
(at $z_a \approx 0.9467$) are marked below. 
\medskip

\noindent\ul{Fig. 2.} --- Observed spectrum of 3C~288.1 (solid histogram) 
and a fit (dotted lines) to the continuum, \ion{O}{6} emission line, 
and several absorption lines. The 
\Lyb\ line at the left has redshift $z_a\approx 0.9467$, while the 
others belong to the strong associated system at $z_a \approx 0.9627$ 
(see Fig. 1 and \S3.1).
\medskip

\noindent\ul{Fig. 3.} --- Spectrum of 3C~288.1 on a log-log scale, 
where $\nu$ is the rest-frame frequency in Hz and the flux $F_{\nu}$ 
has units \fnu . The bold solid line is a broken power law with 
$\alpha = -1.73$ for $\log\nu ({\rm Hz})\ga 15.46$ and 
$\alpha\approx -0.83$ for $\log\nu ({\rm Hz})\la 15.46$. The dotted 
line is just an extension of the low frequency power law segment. 
``LL'' marks the Lyman limit in the $z_a \approx 0.9627$ 
absorption system. 
\medskip

\noindent\ul{Fig. 4.} --- Ionization fractions in optically 
thin clouds that are photoionized at different $U$. The ionizing 
spectrum is a broken power-law with $\alpha_{uvx} =$~$-$1.7 and 
$\alpha_x = -0.9$. The HI fraction 
appears across the top. The curves for the metal ions 
are labeled directly above or below their maximum values, except 
for \ion{Mg}{2} in the lower left of the bottom panel. The sulfur ions 
in the lower panel, and silicon ions in the upper panel, are represented 
by dash-dot curves for clarity. The 
notation is HI = H$^o$, CIV = C$^{+3}$, etc. The bold vertical lines 
connect pairs of $f$(M$_i$) curves at the $U$ value implied by the 
ratio of their column densities. Specifically, the bold lines 
correspond to the ratios \ion{C}{3}/\ion{C}{4}, \ion{O}{4}/\ion{O}{5}, 
\ion{O}{3}/\ion{O}{4}, \ion{O}{5}/\ion{O}{6}, \ion{O}{6}/\ion{Ne}{8}, 
and \ion{Ne}{8}/\ion{Mg}{10} from left to right in the top panel, 
and \ion{S}{4}/\ion{S}{5}, \ion{S}{3}/\ion{S}{4}, 
\ion{N}{3}/\ion{N}{4}, and \ion{N}{4}/\ion{N}{5} in the bottom panel. 
See \S4.3.
\newpage

\plotone{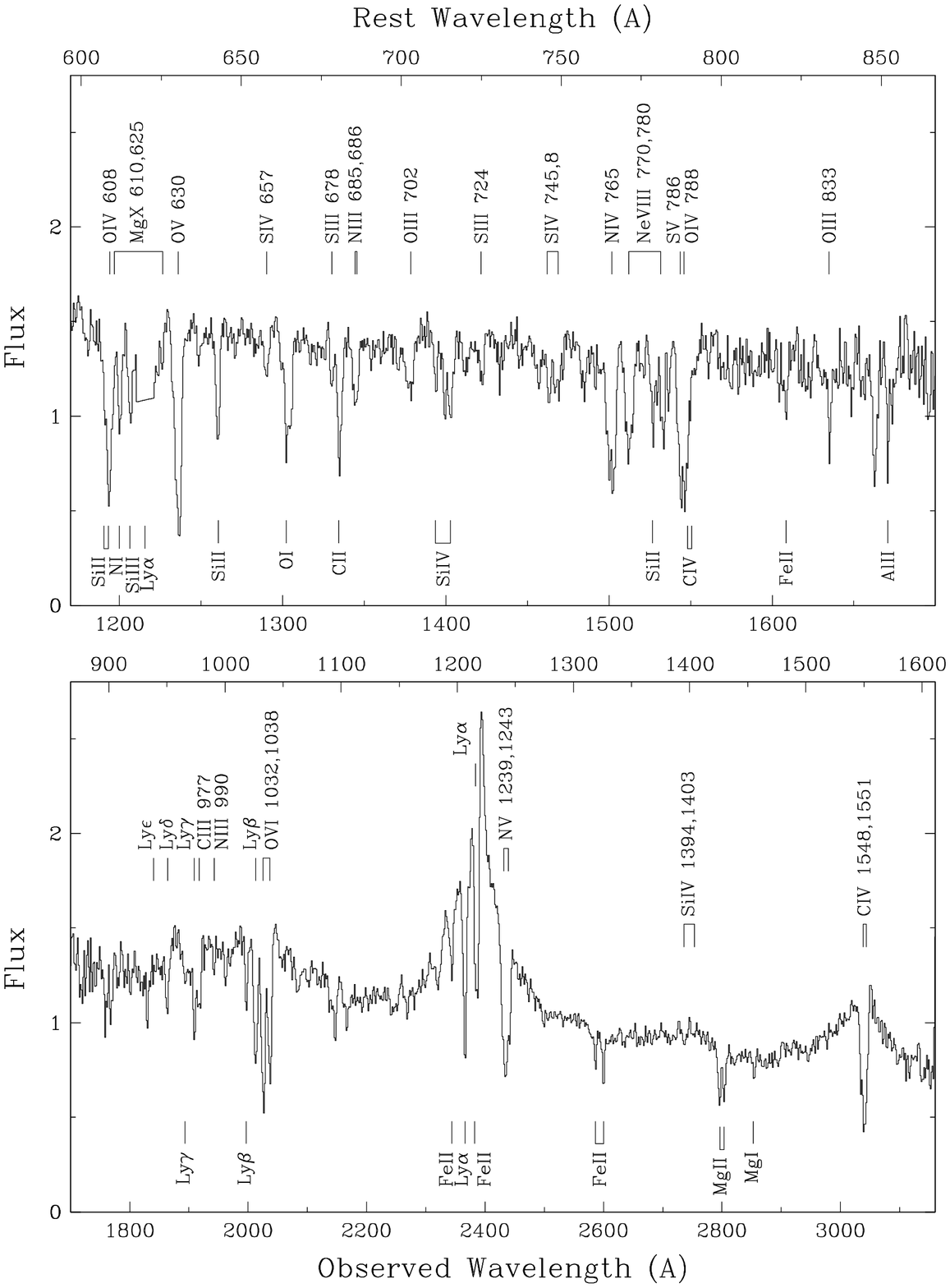}

FIGURE 1.
\newpage

\plotone{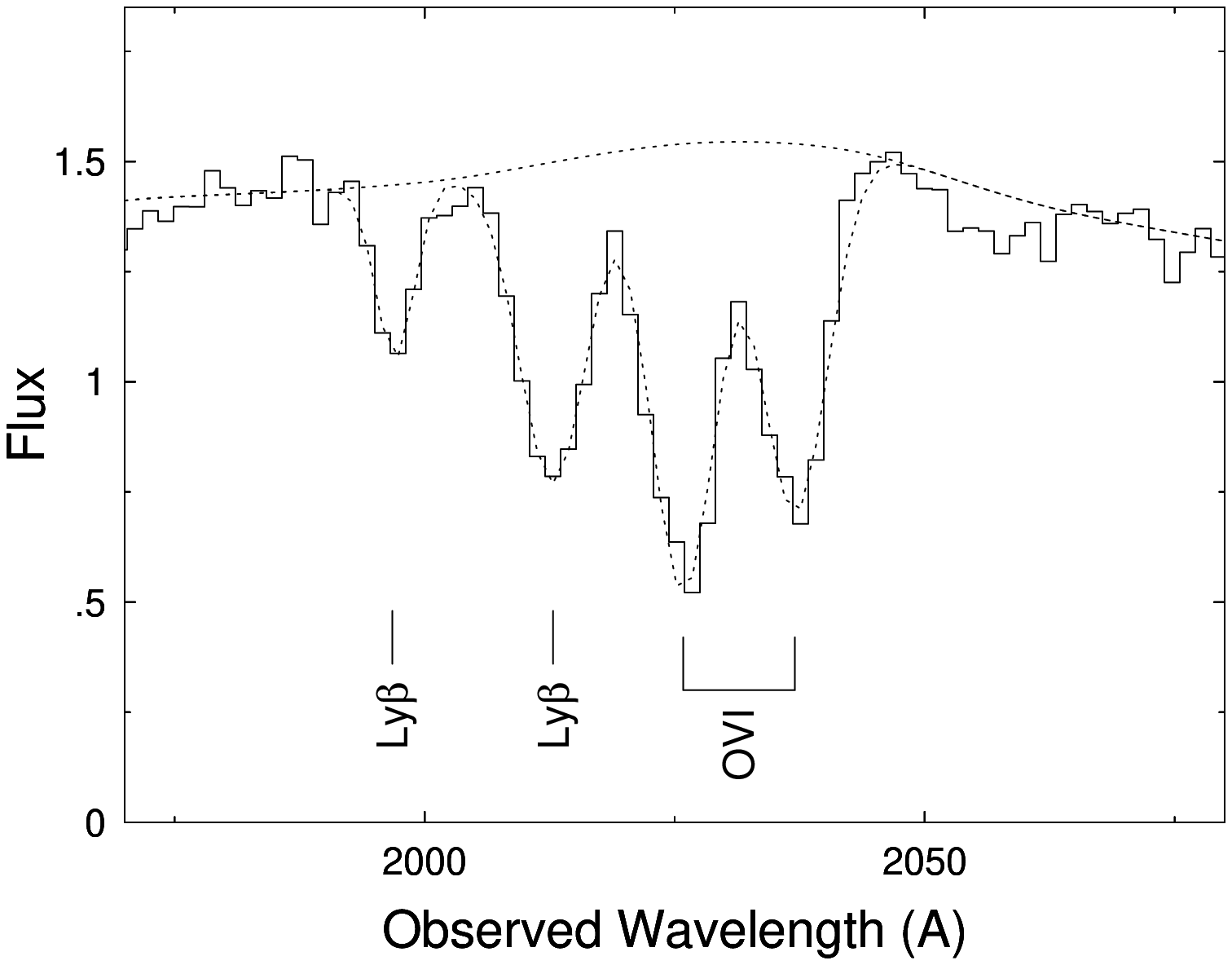}

FIGURE 2.
\newpage

\plotone{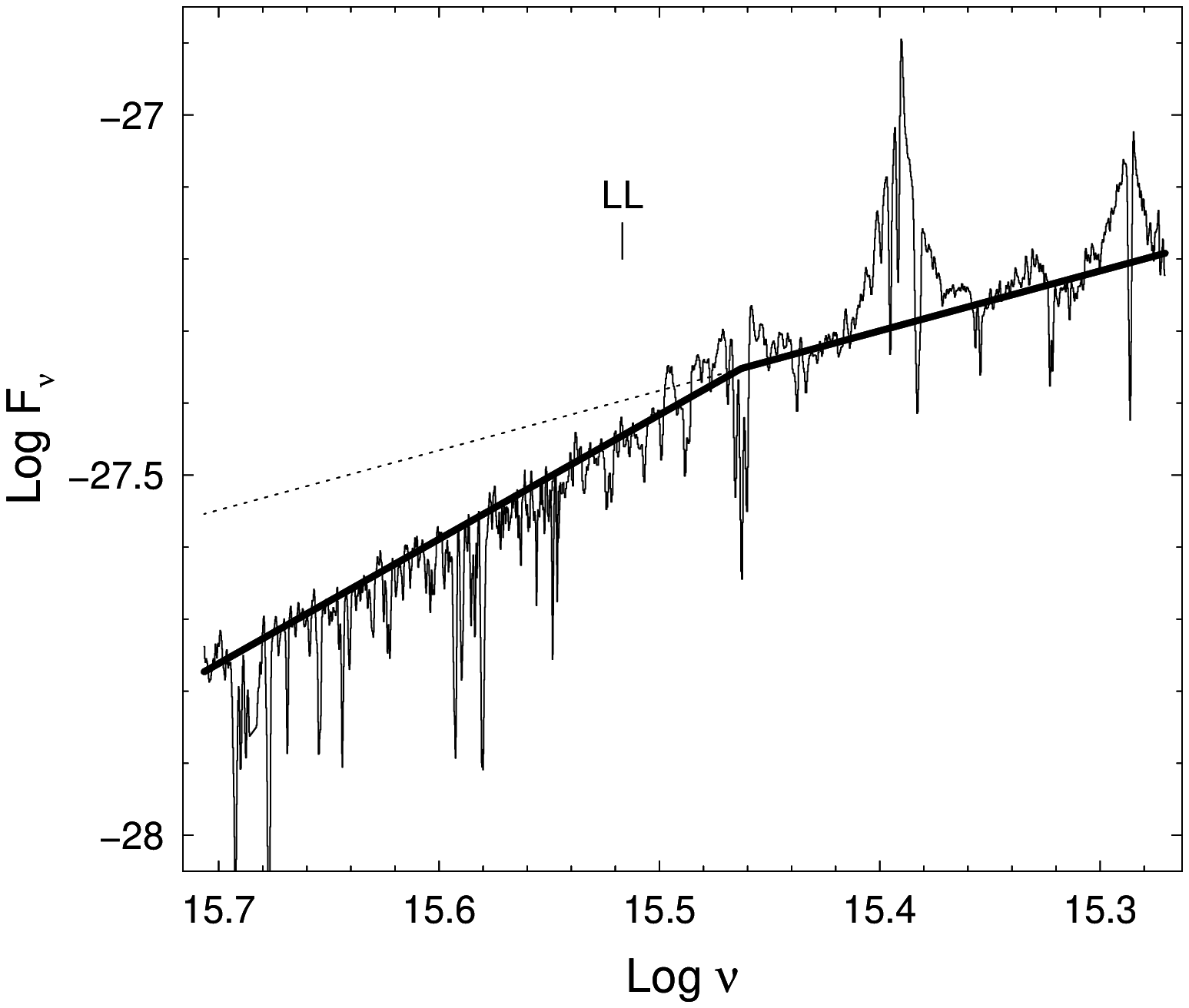}

FIGURE 3.
\newpage

\plotone{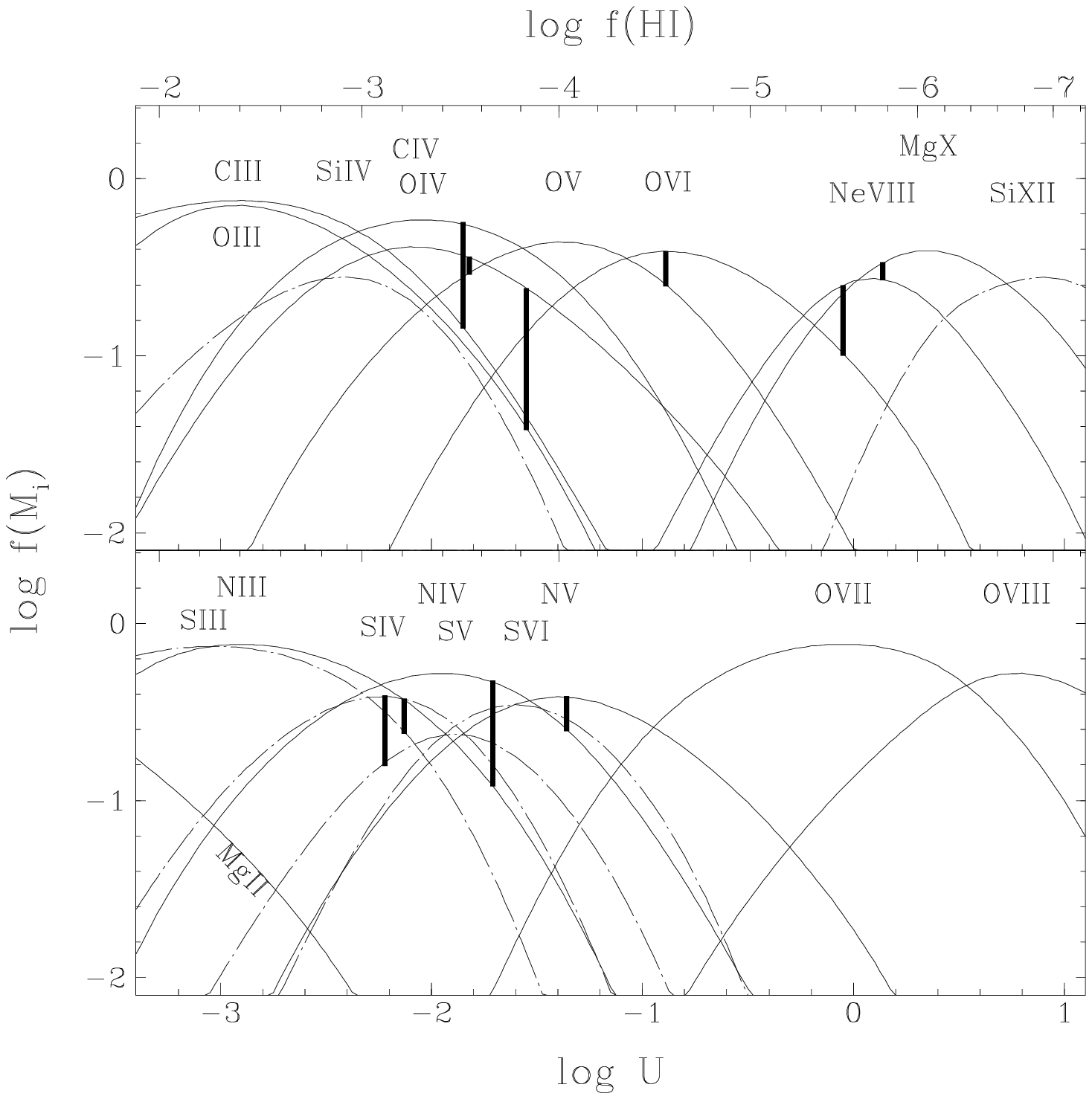}

FIGURE 4.

\end{document}